\documentstyle[aps,preprint,prb,tighten]{revtex}
\begin{document}

\draft
\preprint{MIC-DFM preprint}
\title{Plasmon enhancement of Coulomb drag in double quantum well systems}

\author{Karsten Flensberg$^{1,2}$ and Ben Yu-Kuang Hu$^{1}$}

\address{$\mbox{}^1$Mikroelektronik Centret,\\
Bygning 345 \o st, Danmarks Tekniske Universitet,\\
DK-2800 Lyngby, Denmark}
\address{$\mbox{}^2$Dansk Institut for Fundamental Metrologi,\\
Bygning 307, Anker Engelunds Vej 1,\\
DK-2800 Lyngby, Denmark}

\date{\today}
\maketitle

\begin{abstract}

We derive an expression for the drag rate (i.e., interlayer momentum
transfer rate) for carriers in two coupled two-dimensional
gases to lowest nonvanishing order in the screened interlayer
electron--electron interaction, valid for {\sl arbitrary} intralayer
scattering mechanisms, using the Boltzmann transport equation.
We calculate the drag rate for experimentally relevant parameters, and
show that for moderately high temperatures ($T\gtrsim 0.2 T_F$, where
$T_F$ is the Fermi temperature) the dynamical screening of the interlayer
results in a large enhancement
of the drag rate due to the presence of coupled plasmon modes.
This plasmon enhancement causes the scaled drag rate to have a peak
(i) as a function of temperature at $T \approx 0.5 T_F$, and
(ii) as a function of the ratio of densities of the
carriers in the two layers when their Fermi velocities are equal.
We also show that the drag rate can be
significantly affected by the {\sl intralayer} scattering mechanisms;
in particular, the drag rate changes approximately by a factor of 2
when the dopant layer modulation doped structures are moved in from
400~\AA\ to 100~\AA.

\end{abstract}


\narrowtext
\section{Introduction}
\label{sec:intro}

Coupled quantum wells fabricated by epitaxical growth
make up interesting systems for studying electron--electron
interactions in the low-dimensional systems both experimentally
and theoretically. In coupled wells placed
within tunneling distance, one can for instance study Coulomb gaps
in high magnetic fields.\cite{eise92} For wells separated
so that tunneling can be disregarded, the effect
of the mutual polarization may become important even at
zero magnetic field, and it has been suggested that the interlayer
correlations can drive a Wigner crystallization\cite{swie91} when
the coupled plasmon mode goes soft.
This is however only relevant for low density
systems and has so far not been observed experimentally.
At higher densities, coupled collective modes in multilayer systems
have been seen in inelastic light scattering
experiments.\cite{pinc86,faso86} In this paper, we
discuss another probe of interlayer interactions and the
possibility of observing the coupled
plasmon modes in bilayer systems, namely by the so-called
Coulomb drag effect,\cite{pogr77,pric83}
where a current in one layer drives a current in the
other layer due to the momentum loss caused by interlayer
electron--electron scattering events. The Coulomb drag
effect has recently attracted much experimental
\cite{solo89,gram91,gram93,siva92} and
theoretical attention.\cite{dragtheo,jauh93,zhen93,flen94,swie95}

Normally, the effects of electron--electron collisions only have indirect
consequences for transport properties of single isolated quantum wells,
because they conserve momentum.
The Coulomb drag effect is unique in that it provides
an opportunity to directly measure electron--electron interaction
through a transport measurement where momentum is transferred
from one layer to the other.  Since interlayer interaction
well depend strongly on the many-body effects of the
system it is therefore essential to include screening in the
theoretical understanding of the
measurements.\cite{dragtheo,jauh93,zhen93,flen94,swie95}
This has been realized in much of the theoretical work so
far.  Recently, we have pointed out that at intermediate
temperatures on the scale of the Fermi temperature
the drag effect should in fact be greatly
enhanced by ``anti-screening'' due to coupled plasmon
modes.\cite{flen94}

In this paper, we offer a more detailed study of the predicted
plasmon enhancement effect in zero magnetic field. We base our calculation
on the Random Phase Approximation (RPA) which
is generally believed to be valid for not too low
densities such as e.g. the experimental realization
fabricated by Gramila {\it et al.}\cite{gram91}
However, for low density systems such as the electron-hole
system studied by Sivan {\it et al.},\cite{siva92}
the RPA approximation is not able to account for the
magnitude of the observed drag effect. This discrepancy
between theory and experiment could be an indication of
break-down of RPA due correlation effects.\cite{swie95}
The Sivan {\it et al.} system is however not relevant for the
plasmon enhancement effect studied here due to the
large mismatch between the Fermi velocities,\cite{flen94} as
will be explained below.

The results obtained in this paper are based on the Boltzmann equation.
Quantum effects neglected by the Boltzmann equation can be shown to
be experimentally negligible in the structures we
consider.\cite{flen95a,kame95}
We derive a generalized formula for the drag rate which
is valid for arbitrary intralayer scattering processes,
which reduces to the previously obtained
results\cite{gram91,jauh93,zhen93}
for the case of a constant relaxation time approximation.

The role of intralayer carrier--carrier interactions
has not been studied so far. As mentioned above
the intralayer interactions usually only have indirect
consequences for the transport properties.
However, it turns out that the plasmon part of
the drag effect is very sensitive to the details
of the distribution function. We have solved
for the distribution function including intralayer
electron-electron interaction using parameters
for GaAs-AlGaAs quantum wells and found that:
(1) for high mobility samples where the distant to
the $\delta$-doped impurities is fairly large
the distrubution is well-described by a shifted
Fermi-Dirac distribution function and which allows us
to assume an energy independent momentum relaxation
rate when calculating the drag rates. This is
the situation previously studied by several
authors.\cite{gram91,jauh93,zhen93,flen94}
However, for the case where the dopants is
moved closer to the two-dimensional electron gas
the situation is changed and we
predict and enhancement of the drag rate because
of a larger plasmon contribution in this case.
The reason for this enhancement is that distribution
function becomes skewed toward higher energies and
the overlap with the plasmon branch consequently becomes larger.

The paper is organized as follows. In Sec.\ \ref{sec:dragrate}
we discuss the general formula for the drag rate, and in
Sec.\ \ref{sec:rpa} we study the RPA approximation of the
screening for a coupled two-layer system.  The drag rate for the case of
shifted Fermi-Dirac distributions, which is applicable when the charged
dopants are far from the quantum wells, is studied in Sec.\ \ref{sec:shiftFD}.
In Sec.\ \ref{sec:tauvary}, we study the case when distribtions deviates from
shifted Fermi-Dirac functions,
as when the dopants are close to the quantum wells.
Finally, a summary is given in Sec.\ \ref{sec:disc}.
Technical details are given in a number of appendices.

\section{Boltzman equation calculation of the drag rate}
\label{sec:dragrate}

In a drag experiment, two doped and independently contacted
quantum wells are placed very close to each other.\cite{gram91}
We consider the case typical in experiments where
one applies electric field $E_1$ causing a current of density $J_1$
to flow in layer one.   This sets up an induced electric field $E_2$
in layer two, where the current is set to zero.
(Throughout this paper we denote layer 1 as the driven layer and
layer 2 as the dragged layer.)
The transresistivity tensor $\rho_{21}^{\beta\alpha}$ is then defined as
\begin{equation}
\sum_{\alpha=x,y}\rho_{21}^{\beta\alpha}J_{1,\alpha}  = E_{2,\beta}
\end{equation}
By the Onsager relationship, $\rho_{12}^{\alpha\beta} =
\rho_{21}^{\beta\alpha}$.

In the case of isotropic parabolic bands, where one has a well-defined
effective mass $m$,
one can define a drag rate $\tau_{21}^{-1}$ through
\begin{equation}
\rho_{21}^{\alpha\alpha} \equiv \frac{m_1}{n_1 e^2 \tau_{21}}
\label{deftau21}
\end{equation}
where $n_1$ is the carrier density in layer 1.
Physically, the drag rate is the net average rate of momentum transfered
to each particle in layer 2, per unit drift momentum per particle in
layer 1; i.e.,
\begin{equation}
\frac{1}{\tau_{21}} =
\frac{\overline{(\partial p_2/\partial t)}}{\overline{p}_1}
\end{equation}
where $p_i$ is the momentum per particle in layer $i$, and the
overbar denotes an ensemble average.  Note that by this definition,
the drag rate is not symmetric; i.e.,
$\tau_{12}^{-1} = m_2 n_1 m_1^{-1} n_2^{-1} \tau_{21}^{-1}$,
which is important when layers 1 and 2 are not identical.
We follow the convention of previous authors by denoting $\tau_{21}$ as
$\tau_D$.

The drag rate has previously been derived using a Boltzmann equation
approach\cite{gram91,jauh93} assuming a {\em energy-independent}
intralayer momentum relaxation time, and also using a memory functional
method with a constant relaxation time approximation.  This result
was recently generalized, using the Kubo formalism, to include
energy-dependent impurity scattering rates.\cite{flen95a}
Here we shall further generalize this result to {\em arbitrary}
intralayer scattering mechanism
within the linear response Boltzmann equation description.

\subsection{Boltzmann equation for coupled quantum wells}

We define the function $\psi(\mbox{\boldmath{$k$}})$ which is related to the
deviation of the distribution function from equilibrium by
\begin{eqnarray}
\delta\! f(\mbox{\boldmath{$k$}}) \equiv
f(\mbox{\boldmath{$k$}})-f^0(k) &\equiv& f^0(k)[1-f^0(k)]
\psi(\mbox{\boldmath{$k$}}) \nonumber\\
&=& -k_B T \left(\frac{\partial f^0(\mbox{\boldmath{$k$}})}
{\partial\varepsilon_k}\right)\;
\psi(\mbox{\boldmath{$k$}}),
\label{definepsi}
\end{eqnarray}
where $f^0(\mbox{\boldmath{$k$}})$ is the equilibrium Fermi-Dirac distribution
function.  With this definition, the linearized interlayer
electron--electron collision term is\cite{jauh93,smith}
\widetext
\begin{eqnarray}
S[\psi_1,\psi_2](\mbox{\boldmath{$k$}}_2) &=&
2\int {d\mbox{\boldmath{$k$}}_1\over (2\pi)^2 }
\int {d\mbox{\boldmath{$q$}}\over (2\pi)^2}\
w(\mbox{\boldmath{$q$}},\varepsilon_{\mbox{\boldmath{$k$}}_1+\mbox{\boldmath{$q$}}} -
\varepsilon_{\mbox{\boldmath{$k$}}_1})\nonumber\\
&& f^0_1(\mbox{\boldmath{$k$}}_1) f^0_2(\mbox{\boldmath{$k$}}_2)
[1-f^0_1(\mbox{\boldmath{$k$}}_1+\mbox{\boldmath{$q$}})]
[1-f^0_2(\mbox{\boldmath{$k$}}_2-\mbox{\boldmath{$q$}})] \nonumber\\
&&\times \left[\psi_1(\mbox{\boldmath{$k$}}_1) +
\psi_2(\mbox{\boldmath{$k$}}_2) - \psi_1(\mbox{\boldmath{$k$}}_1 +
\mbox{\boldmath{$q$}})
-\psi_2(\mbox{\boldmath{$k$}}_2 - \mbox{\boldmath{$q$}})\right]\nonumber\\
&&\delta\left(\varepsilon_{\mbox{\boldmath{$k$}}_1} +
\varepsilon_{\mbox{\boldmath{$k$}}_2}
-\varepsilon_{\mbox{\boldmath{$k$}}_1 + \mbox{\boldmath{$q$}}} -
\varepsilon_{\mbox{\boldmath{$k$}}_2 - \mbox{\boldmath{$q$}}}\right).
\label{linintercoll}
\end{eqnarray}
\narrowtext
Here $w(q,\omega)$ is the probability of a particle scattering with
change of momentum and energy of $\hbar\mbox{\boldmath{$q$}}$ and
$\hbar\omega$, respectively.
This rate is usually taken to be the Born approximation result
$w(q,\omega) = 2\pi\hbar^{-1} |W_{12}(q,\omega)|^2,$
where $W_{12}(q,\omega)$ is the dynamically screened interlayer
Coulomb interaction matrix element.\cite{bandcompl}

We assume that the interlayer interactions are weak,
so that these interactions are only kept to lowest
order.  Then, within linear response to an external driving field
$E_1$, the coupled Boltzmann equations for the system reads
\begin{mathletters}
\begin{eqnarray}
e_1
\mbox{\boldmath{$E$}}_1\cdot\mbox{\boldmath{$v$}}\;\Bigl({\partial{f^0_1}\over\partial{\varepsilon}}\Bigr) &=& -\hat
H_1[\psi_1](\mbox{\boldmath{$k$}}_1),\label{coupboltz1}\\
e_2
\mbox{\boldmath{$E$}}_2\cdot\mbox{\boldmath{$v$}}\;\Bigl({\partial{f^0_2}\over\partial{\varepsilon}}\Bigr) &=&
S[\psi_1,\psi_2=0](\mbox{\boldmath{$k$}}_2) - \hat H_2
\left[\psi_2\right](\mbox{\boldmath{$k$}}_2).
\label{coupboltz2}
\end{eqnarray}
\end{mathletters}
where $e_i$ is the carrier charge in layer $i$, $\hat H_i$ is the
negative of the linearized {\em intra}layer collision operator.\cite{smith}
We have used the assumption of weak interlayer interaction
to neglect the interlayer electron--electron collision term
in Eq.\ (\ref{coupboltz1}) and to set $\psi_2(\mbox{\boldmath{$k$}}_2) = 0$ in
the interlayer collision term $S$ in Eq.\ (\ref{coupboltz2}),
as these terms are higher order in the interlayer interaction than
the other terms in their respective equations. For convenience,
hereafter we shall refer to $S[\psi_1,\psi_2=0](\mbox{\boldmath{$k$}}_2)$
simply
as $S(\mbox{\boldmath{$k$}}_2)$.

The formal solutions for $\psi_1(\mbox{\boldmath{$k$}}_1)$ and
$\psi_2(\mbox{\boldmath{$k$}}_2)$ for
$\mbox{\boldmath{$E$}}$ fields in the $x$-direction are
\begin{mathletters}
\begin{eqnarray}
\psi_1(\mbox{\boldmath{$k$}}_1) &=&  -e_1 E_1
\hat H_1^{-1} \left[v_{x,1}\,
\Bigl({\partial{f^0_1}\over\partial{\varepsilon}}\Bigr)
\right](\mbox{\boldmath{$k$}}_1)\label{formalpsi1}\\
\psi_2 (\mbox{\boldmath{$k$}}_2) &=& e_2 E_2
\hat H_2^{-1}\left[v_{x,2}
\Bigl({\partial{f^0_2}\over\partial{\varepsilon}}\Bigr)\right](\mbox{\boldmath{$k$}}_2)
+ \hat H_2^{-1}[S](\mbox{\boldmath{$k$}}_2).
\label{formalpsi2}
\end{eqnarray}
\end{mathletters}
We assume the electric field in layer 2 is adjusted so that the
current in layer 2 is zero, i.e.,
\begin{equation}
j_{2,x} = -2 e_2 k_B T \int \frac{d\mbox{\boldmath{$k$}}_2}
{(2\pi)^2} v_{2,x}
\Bigl({\partial{f^0_2}\over\partial{\varepsilon}}\Bigr)
\psi_2(\mbox{\boldmath{$k$}}_2) =0,
\label{zerocur}
\end{equation}
which implies, by substitution of Eq. (\ref{formalpsi2})
into Eq. (\ref{zerocur}),
\widetext
\begin{eqnarray}
2 k_B T e_2^2 E_2 \int \frac{d\mbox{\boldmath{$k$}}_2}{(2\pi)^2}\
v_{2,x} \Bigl({\partial{f^0_2}\over\partial{\varepsilon}}\Bigr)\ \hat H_2^{-1}
\left[v_{2,x}
\Bigl({\partial{f^0_2}\over\partial{\varepsilon}}\Bigr)\right](\mbox{\boldmath{$k$}}_2)
\equiv&& n_2 e_2 \mu_{t,2} E_2\nonumber\\
= - 2 e_2 k_B T \int \frac{d\mbox{\boldmath{$k$}}_2}{(2\pi)^2}&&
v_{x,2}\Bigl({\partial{f^0_2}\over\partial{\varepsilon}}\Bigr) \hat
H_2^{-1}[S](\mbox{\boldmath{$k$}}_2).
\label{balance}
\end{eqnarray}
The equivalence in Eq. (\ref{balance})
comes from Eq.\ (\ref{jbeta}) in Appendix \ref{app:boltz},
and the $\mu_{t,2}$ is the mobility of layer 2
in the absence of interlayer coupling.

Substituting Eq. (\ref{linintercoll}) (with $\psi_2(\mbox{\boldmath{$k$}}_2) =
0$)
into Eq.  (\ref{balance}), and using the identities
\begin{mathletters}
\begin{eqnarray}
\delta\left(\varepsilon_{\mbox{\boldmath{$k$}}_1} +
\varepsilon_{\mbox{\boldmath{$k$}}_2}
-\varepsilon_{\mbox{\boldmath{$k$}}_1 + \mbox{\boldmath{$q$}}} -
\varepsilon_{\mbox{\boldmath{$k$}}_2 -
\mbox{\boldmath{$q$}}}\right)
&=& \hbar\int_{-\infty}^\infty d\omega \
\delta(\varepsilon_{\mbox{\boldmath{$k$}}_1}
-\varepsilon_{\mbox{\boldmath{$k$}}_1 + \mbox{\boldmath{$q$}}} - \hbar\omega)
\delta(\varepsilon_{\mbox{\boldmath{$k$}}_2}
- \varepsilon_{\mbox{\boldmath{$k$}}_2 -
\mbox{\boldmath{$q$}}}+ \hbar\omega);\\
f^0(\varepsilon_1)[1 - f^0(\varepsilon_2)] &=&
[f^0(\varepsilon_2) - f^0(\varepsilon_1)] n_B(\varepsilon_1-\varepsilon_2),
\end{eqnarray}
\end{mathletters}
one obtains
\begin{eqnarray}
n_2 e_2 \mu_{t,2} E_2
&=& - 4 e_2 \hbar k_B T
\int \frac{d\mbox{\boldmath{$q$}}}{(2\pi)^2}\
\int_{-\infty}^\infty d\omega\
w(\mbox{\boldmath{$q$}},\omega)\, n_B(\hbar\omega)\, n_B(-\hbar\omega)
\nonumber\\
&& \times\Biggl[\int \frac{d\mbox{\boldmath{$k$}}_1}{(2\pi)^2}\
[\psi_1(\mbox{\boldmath{$k$}}_1) -
\psi_1(\mbox{\boldmath{$k$}}_1 + \mbox{\boldmath{$q$}})]
[f^0_1(\mbox{\boldmath{$k$}}_1)-f^0_1(\mbox{\boldmath{$k$}}_1+\mbox{\boldmath{$q$}})]
\delta(\varepsilon_{\mbox{\boldmath{$k$}}_1} -
\varepsilon_{\mbox{\boldmath{$k$}}_1+
\mbox{\boldmath{$q$}}} - \hbar\omega)\Biggr]
\nonumber\\
&& \times\Biggl\{\int \frac{d\mbox{\boldmath{$k$}}_2}{(2\pi)^2}\
v_{x,2} \Bigl({\partial{f^0_2}\over\partial{\varepsilon}}\Bigr)
\hat
H_2^{-1}\left[\left\{f_2^0(\mbox{\boldmath{$k$}}_2)-f_2^0(\mbox{\boldmath{$k$}}_2 - \mbox{\boldmath{$q$}})\right\}
\delta(\varepsilon_{\mbox{\boldmath{$k$}}_2} -
\varepsilon_{\mbox{\boldmath{$k$}}_2 -
\mbox{\boldmath{$q$}}} + \hbar\omega)\right]\Biggr\}.
\label{gruesome}
\end{eqnarray}

To write Eq.\ (\ref{gruesome}) in a more tractable form,
we {\em define} the transport relaxation time of layer $i$,
$\tau_i(\mbox{\boldmath{$k$}}_i)$, for arbitrary intralayer scattering by
\begin{equation}
\psi_i(\mbox{\boldmath{$k$}}_i) =
-e\mbox{\boldmath{$E$}}_i\cdot\hat H_i^{-1}\left[\mbox{\boldmath{$v$}}_i
\left(\frac{\partial f^0}
{\partial\varepsilon}\right)\right](\mbox{\boldmath{$k$}}_i)
\equiv \frac{e_i
\mbox{\boldmath{$E$}}_i\cdot\mbox{\boldmath{$v$}}(\mbox{\boldmath{$k$}}_i)\,
\tau_i(\mbox{\boldmath{$k$}}_i)}{k_B T},
\label{definetau}
\end{equation}
in analogy with the case of impurity-dominated scattering
(see Appendix \ref{app:boltz}).
Then, from Eq.\ (\ref{formalpsi1}) and the definition
in Eq.\ (\ref{definetau}), the term in the large square brackets
in Eq.\ (\ref{gruesome})
(i.e., involving integration over {\boldmath $k$}$_1$)
is equal to
\begin{eqnarray}
\Biggl[ \cdots \Biggr]\ \mbox{in Eq. (\protect\ref{gruesome})}
&=&\int \frac{d\mbox{\boldmath{$k$}}_1}{(2\pi)^2}\
[\psi_1(\mbox{\boldmath{$k$}}_1) -
\psi_1(\mbox{\boldmath{$k$}}_1 + \mbox{\boldmath{$q$}})]
[f^0_1(\mbox{\boldmath{$k$}}_1)-f^0_1(\mbox{\boldmath{$k$}}_1+\mbox{\boldmath{$q$}})]
\delta(\varepsilon_{\mbox{\boldmath{$k$}}_1} -
\varepsilon_{\mbox{\boldmath{$k$}}_1+
\mbox{\boldmath{$q$}}} - \hbar\omega)\nonumber\\
&=&  -\frac{e_1 E_1}{k_B T}
\int \frac{d\mbox{\boldmath{$k$}}_1}{(2\pi)^2}\ [
v_{x,1}(\mbox{\boldmath{$k$}}_1 +
\mbox{\boldmath{$q$}})\tau_1(\mbox{\boldmath{$k$}}_1 +
\mbox{\boldmath{$q$}})-v_{x,1}(\mbox{\boldmath{$k$}}_1)\tau_1(\mbox{\boldmath{$k$}}_1)]
\nonumber \\
&&\times
[f^0_1(\mbox{\boldmath{$k$}}_1)-f^0_1(\mbox{\boldmath{$k$}}_1+\mbox{\boldmath{$q$}})]
\delta(\varepsilon_{\mbox{\boldmath{$k$}}_1} -
\varepsilon_{\mbox{\boldmath{$k$}}_1+ \mbox{\boldmath{$q$}}} - \hbar\omega).
\label{int1}
\end{eqnarray}
Note that this quantity is {\em odd} with respect
to $\omega$.\cite{oddness}

Furthermore, using the fact that $\hat H_2^{-1}$ is Hermitian
(see Appendix \ref{app:boltz}) and Eq.\ (\ref{definetau}),
the term in Eq.\ (\ref{gruesome}) in the large curly braces
(i.e., involving integration over $\mbox{\boldmath{$k$}}_2$) can be rewritten
as
\begin{eqnarray}
\Biggl\{\cdots\Biggr\}\ \mbox{in Eq. (\protect{\ref{gruesome}})}
&=&\int \frac{d\mbox{\boldmath{$k$}}_2}{(2\pi)^2}\
\left\{f_2^0(\mbox{\boldmath{$k$}}_2)-f_2^0(\mbox{\boldmath{$k$}}_2 -
\mbox{\boldmath{$q$}})\right\}
\delta(\varepsilon_{\mbox{\boldmath{$k$}}_2} -
\varepsilon_{\mbox{\boldmath{$k$}}_2 -
\mbox{\boldmath{$q$}}} + \hbar\omega) \nonumber\\
&& \times\hat H_2^{-1}[v_{x,2}
\Bigl({\partial{f^0_2}\over\partial{\varepsilon}}\Bigr)](\mbox{\boldmath{$k$}}_2)\nonumber\\
&=& -\frac{1}{k_B T}\int \frac{d\mbox{\boldmath{$k$}}_2}{(2\pi)^2}\
\left\{f_2^0(\mbox{\boldmath{$k$}}_{2}) -f_2^0(\mbox{\boldmath{$k$}}_2 -
\mbox{\boldmath{$q$}})\right\}
\delta(\varepsilon_{\mbox{\boldmath{$k$}}_2} -
\varepsilon_{\mbox{\boldmath{$k$}}_2 -
\mbox{\boldmath{$q$}}} + \hbar\omega)
v_{x,2}(\mbox{\boldmath{$k$}}) \tau_2(k)\nonumber\\
&=& \frac{1}{2 k_B T} \int \frac{d\mbox{\boldmath{$k$}}_2'}{(2\pi)^2}\
\left\{f_2^0(\mbox{\boldmath{$k$}}_2')
-f_2^0(\mbox{\boldmath{$k$}}_2'+\mbox{\boldmath{$q$}})\right\}\nonumber\\
&&\times\Bigl[\delta(\varepsilon_{\mbox{\boldmath{$k$}}_2'}
-  \varepsilon_{\mbox{\boldmath{$k$}}_2' + \mbox{\boldmath{$q$}} }
- \hbar\omega) v_{x,2}(\mbox{\boldmath{$k$}}_2'+\mbox{\boldmath{$q$}})
\tau_2(\mbox{\boldmath{$k$}}_2'+\mbox{\boldmath{$q$}}) \nonumber\\
&& \ \ + \delta(\varepsilon_{\mbox{\boldmath{$k$}}_2'} -
\varepsilon_{\mbox{\boldmath{$k$}}_2' +
\mbox{\boldmath{$q$}}} +  \hbar\omega) v_{x,2}(\mbox{\boldmath{$k$}}_2')
\tau_2(\mbox{\boldmath{$k$}}_2')\Bigr].
\label{int2}
\end{eqnarray}
For the last equality in Eq.\ (\ref{int2}),
we have symmetrized by dividing into two equal
parts and performing the variable changes: $\mbox{\boldmath{$k$}}_2 =
 \mbox{\boldmath{$k$}}_2' + \mbox{\boldmath{$q$}}$ in the first term and
$\mbox{\boldmath{$k$}}_2 = - \mbox{\boldmath{$k$}}_2'$
in the second term and using inversion symmetry in
$\mbox{\boldmath{$k$}}$-space.
Inserting Eqs.\ (\ref{int1}) and (\ref{int2}) into
Eq.\  (\ref{gruesome}), and performing a
variable change $\omega\rightarrow -\omega$
in the second term in the square brackets of Eq.\ (\ref{int2})
(which gives an overall minus sign because of the oddness of
Eq.\ (\ref{int1}) with respect to $\omega$)
gives the following symmetric form for the transresistivity tensor
\begin{equation}
\rho_{21}^{\beta\alpha} =  \frac{E_2}{n_1 e_1 \mu_{t,1} E_1}
= \frac{\hbar^2}{2\pi e_1 e_2 n_1 n_2 k_B T}
\int \frac{d\mbox{\boldmath{$q$}}}{(2\pi)^2} \int_0^\infty d\omega\,
\frac{|W_{12}(q,\omega)|^2 F_1^\alpha(\mbox{\boldmath{$q$}},\omega)]
F^\beta_2(\mbox{\boldmath{$q$}},\omega)}{\sinh^2(\hbar\omega/[2k_B T])}.
\label{rho21}
\end{equation}
Here, we have used $4 n_B(\hbar\omega)n_B(-\hbar\omega) =
-\sinh^{-2}(\hbar\omega\beta/2)$
and we have defined the function
\begin{eqnarray}
F^\alpha (\mbox{\boldmath{$q$}},\omega) &=&
\frac{2\pi e}{\hbar\mu_t}
\int \frac{d\mbox{\boldmath{$k$}}}{(2\pi)^2}
[f^0(\mbox{\boldmath{$k$}})-f^0(\mbox{\boldmath{$k$}}+\mbox{\boldmath{$q$}})]\;\delta(\varepsilon(\mbox{\boldmath{$k$}})
- \varepsilon(\mbox{\boldmath{$k$}} + \mbox{\boldmath{$q$}}) -
\hbar\omega)\nonumber\\
&&\times[v^\alpha(\mbox{\boldmath{$k$}}+\mbox{\boldmath{$q$}})\tau(\mbox{\boldmath{$k$}}+\mbox{\boldmath{$q$}})-v^\alpha(\mbox{\boldmath{$k$}})\tau(\mbox{\boldmath{$k$}})].
\label{defF}
\end{eqnarray}
As required by the Onsager relation,
$\rho_{21}^{\beta\alpha} = \rho_{21}^{\alpha\beta}$ in Eq. (\ref{rho21}).

\subsection{Drag rate for isotropic parabolic bands}

We concentrate on the case where the bands are isotropic and parabolic,
as in the case of the electrons or low-energy holes in GaAs, generally
the material of choice so far in performing drag experiments.
In this case, when one has a well-defined mass, the mobility can be
written in terms of the transport time
\begin{equation}
\mu_t = \frac{e\tau_{\mathrm{tr}}}{m}.
\end{equation}
This transport time is the time that enters the
conductivity, $\sigma=ne^2\tau_{\mathrm{tr}}/m$.
One can write $F^\alpha(\mbox{\boldmath{$q$}},\omega) = q^\alpha Y(q,\omega)$,
where
\begin{eqnarray}
Y_i(q,\omega) &=& \frac{2\pi}{q \tau_{\mathrm{tr},i}}
\left|\int \frac{d\mbox{\boldmath{$k$}}}{(2\pi)^2}
\left\{f_i(\varepsilon(\mbox{\boldmath{$k$}}))-f_i(\varepsilon(\mbox{\boldmath{$k+q$}}))\right\}
\right.\nonumber\\
&&
\times\phantom{\int}
\left\{\mbox{\boldmath{$k$}}\tau_i(\varepsilon(\mbox{\boldmath{$k$}}))-
(\mbox{\boldmath{$k+q$}})\tau_i(\varepsilon(\mbox{\boldmath{$k+q$}}))\right\}
\nonumber\\
&&
\times\left.\phantom{\int}
\delta(\varepsilon(\mbox{\boldmath{$k$}})-\varepsilon(\mbox{\boldmath{$k+q$}})-\omega)
\right|.
\label{Fdef}
\end{eqnarray}
Hence, using Eq. (\ref{deftau21}) and Eq. (\ref{rho21}),
the drag-rate for isotropic parabolic bands can be written as
\begin{eqnarray}
\tau_D^{-1} \equiv \tau_{21}^{-1} &=& \frac{\hbar^2}{8m_1 n_2 k_B T
\pi^2} \int_0^\infty dq\ q^3 \int_0^\infty d\omega\, \nonumber \\
&&\times\frac{|W_{12}(q,\omega)|^2
Y_1(q,\omega) Y_2(q,\omega)}{\sinh^2(\hbar\omega/[2k_B T])}.
\label{tauD}
\end{eqnarray}

When $\tau_i(\mbox{\boldmath{$k$}})$ is independent of $\mbox{\boldmath{$k$}}$
the function $Y$
reduces to the usual imaginary part of the polarization function,
Im$[\chi]$, and the drag rate formula in Eq.\ (\ref{tauD}) reduces
the result of Refs.\ \onlinecite{gram91}, \onlinecite{siva92} and
\onlinecite{jauh93}.
{\em We emphasize that results in Eq.\ (\ref{rho21})
and Eq. (\ref{tauD}) hold for arbitrary intralayer scattering mechanism.}
However, in order to compute the drag rate for a given
system one must obtain $\tau(k)$ by first solving for the
linear-response distribution
function for each layer in the absence of interlayer coupling.

In this paper, we study the drag effect at intermediate
temperatures where phonon scattering provides an important
contribution to the scattering rates and hence one may
expect that the solution of the Boltmann equation does
give a non-constant $\tau_i(\mbox{\boldmath{$k$}})$. On the other hand,
electron-electron interactions tend to pull the
distribution function back to a shifted Fermi-Dirac
distribution.
We have performed a numerical calculation
of the distribution function in GaAs based quantum wells
with dopants placed approximately 700~\AA\ away, as in previous
experiments (e.g., in Refs. \onlinecite{gram91} and \onlinecite{siva92})
and found that the distribution functions
in these cases can, to a very good approximation, be described by a
constant scattering time.
Details are given elsewhere.\cite{sanb95,hu95}
The characteristic time scale for electron--electron
scattering is given by\cite{2dscat}
$\tau_{e-e} \approx (h/E_F)\;(E_F/k_B T)^2$.
For GaAs doped at $1.5 \times 10^{11}$ cm$^{-2}$ at 10 K, $\tau_{e-e}
\approx 10^{-11}$ s, as compared to the impurity and phonon scattering times
which are typically of order $\tau_{\mathrm{imp,ph}}\sim 10^{-9}$ s.
The relation $\tau_{e-e} \ll \tau_{\mathrm{imp,ph}}$ explains why the shifted
Fermi-Dirac distribution
function is a good approximation to the actual
solution of the Boltzmann equation.

Therefore, in section \ref{sec:shiftFD} we shall
restrict ourselves to the case where the $Y$-function
in Eq.\ (\ref{Fdef}) can be replaced by the imaginary part of
the polarizability,\cite{gram91,siva92,jauh93}
\begin{eqnarray}
\tau_D^{-1} &=& \frac{\hbar^2}{8m_1 n_2 k_B T
\pi^2} \int_0^\infty dq\ q^3 \int_0^\infty d\omega\, \nonumber \\
&&\times\frac{|W_{12}(q,\omega)|^2
\mathrm{Im}\,\chi_1(q,\omega)
\mathrm{Im}\,\chi_2(q,\omega)}{\sinh^2(\hbar\omega/[2k_B T])}.
\label{dragrate}
\end{eqnarray}
However, we show in section \ref{sec:tauvary} that in the case
when the impurities in the modulation doped samples are moved
in closer to the quantum wells, the momentum relaxation times
$\tau(\mbox{\boldmath{$k$}})$ become significantly dependent on
$\mbox{\boldmath{$k$}}$, and
there are fairly large differences between the drag rates evaluated
by Eq. (\ref{tauD}) and Eq. (\ref{dragrate}).

\section{Coupled plasmon modes in the random phase approximation}

\label{sec:rpa}

To calculate the drag rate, one needs to evaluate the
dynamically screened Coulomb interaction $W(q,\omega)$,
which we do using the RPA.
The RPA equations for the coupled two layer system read\cite{hu93}
\begin{equation}
\left(
\begin{array}{c}
U_{11}\\
U_{21}
\end{array}\right) =
\left(
\begin{array}{c}
V_{11}\\
V_{21}
\end{array}\right) +
\left(
\begin{array}{cc}
V_{11} \chi_1 & V_{12}\chi_2\\
V_{21}\chi_1 & V_{22} \chi_2
\end{array}\right)
\left(
\begin{array}{c}
U_{11}\\
U_{21}
\end{array}\right),\label{RPAequation}
\end{equation}
where the matrix $V_{ij}$ defines
the unscreened Coulomb interaction given by
$V_{ij}=F_{ij}V(q)$.  Here,
$V(q)=2\pi e^2/(\epsilon_0q)
=2\pi \epsilon_Fq_{TF}/(k_F^2q)$ and
$F_{ij}$ are the form factors, for which we use the form
for square wells (see e.g. Ref.\ \onlinecite{jauh93}),
\begin{mathletters}
\begin{eqnarray}
F_{ii}&=& \frac{3 x + 8\pi^2/x}{x^2+4\pi^2}-\frac{32\pi^2(1-\exp(-x))}
{x^2(x^2+4\pi^2)^2},\\
F_{12} &=& \frac{64\pi^4 \sinh^2(x/2)}{x^2(x^2+4\pi^2)^2}e^{-qd}.
\label{formfactor}
\end{eqnarray}
\end{mathletters}
\noindent where $x=qL$, $L$ is the width of the quantum wells
(equal widths are assumed) and $d$ is the center-to-center well
separation.
The solution for the screened interlayer interaction thus becomes
\begin{equation}\label{U12}
U_{12}(q,\omega) = \frac{V_{12}(q)}{\epsilon(q,\omega)},
\end{equation}
where
\begin{eqnarray}
\epsilon(q,\omega)&=&[1-V_{11}(q)\chi_1(q,\omega)][1-V_{22}(q)\chi_2(q,\omega)]
\nonumber \\
&& -[V_{12}(q)]^2\chi_1(q,\omega)\chi_2(q,\omega)  \label{epsilon}
\end{eqnarray}

The collective modes of the coupled electron gas are given
by the zeros of the dielectric function.
There are two such modes, one where the electron densities
in the two layers oscillate in phase, which we call
the optic mode, and one where the oscillations are
out of phase, referred to as the acoustic mode.\cite{dass81}
At zero temperature $U_{12}(q,\omega)$ has two poles on the real
$\omega$-axis, giving $\delta$-function peaks.
At finite temperatures these poles move off the real
$\omega$ axis, which implies that they will gain finite widths.
In calculating the plasmon dispersions at finite $T$, we search
for the solutions of Re$[\epsilon(q,\omega(q))] = 0$; so long as
the damping is small and the poles are not too far from the real
axis, this criterion accurately gives the position of the pole.

While there are simple analytic expressions for Im[$\chi(q,\omega)$]
and Re[$\chi(q,\omega)$] at $T=0$, none exist at finite temperatures.
Since it is necessary to evaluate the $\chi$ at finite temperature
(using the $T=0$ expressions for $\chi$ do {\em not} give the required
plasmon enhancement, as explained later), we have developed an efficient
way of calculating the finite-temperature $\chi$, which is elucidated
in App.\ \ref{app:X}.

In order to gain insight into the structure of the collective mode
dispersions, we take in the following simplifying limit which
allows us to obtain an analytic solution of the dispersion relation of
the the poles at $T=0$.

\subsection{The thin layer or small-$q$ limit of $\epsilon(q,\omega)$}
\label{sec:ana}

In the case when $qL \ll 1$, one can to a good approximation set
$x = qL = 0$ in Eq. (\ref{formfactor}) and
the form factors reduce simply to $F_{11}=1$ and $F_{12}=\exp(-qd)$.
Then, the dielectric function for two identical layers
can be written as
\begin{equation}\label{eps0}
\epsilon(q,\omega) = (1-e^{-2qd})[\chi(q,\omega)-\chi_+(q)]
[\chi(q,\omega)-\chi_-(q)]\{V(q)\}^2,
\end{equation}
where
\begin{mathletters}
\begin{eqnarray}
\chi_\pm(q) &=& \frac{c_\pm(q)}{V(q)};\\
c_\pm(q) &=& \frac{1}{1\pm e^{-qd}}.
\end{eqnarray}
\end{mathletters}
{}From this it is clear that the dispersion of the two collective
modes are given by the conditions $\chi(q,\omega_\pm(q))=\chi_\pm(q)$.
The zero temperature real part of $\chi$ is given by\cite{ster67}
\begin{equation}
V(q)\chi'(q,\omega) = \frac{q_{TF}k_{F}}{q^2}
\left(\frac{q}{k_{F}}-\sqrt{a_+^2-1}+\sqrt{a_-^2-1}\right),
\end{equation}
where we have defined $a_\pm = \omega/v_{F}q\pm q/2k_{F}$,
and used the notation $\chi = \chi'+i\chi''$.
We then find for the plasmon solutions
\begin{mathletters}
\begin{eqnarray}\label{plasdis}
\omega_\pm(q)&=& qv_F\frac{b_\pm\sqrt{1+4/A_\pm}}{2},\\
b_\pm &=& c_\pm \frac{q^2}{q_{TF}k_{F}}
+\frac{q}{k_F},\\
A_\pm &=& 2 c_\pm \frac{q^3}{q_{TF} k_{F}^2}
\left(1+\frac{qc_\pm}{2q_{TF}}\right).
\end{eqnarray}
\end{mathletters}
\noindent The small-$q$ behaviors of the plasmon solutions are
\begin{mathletters}\label{plasdisapp}
\begin{eqnarray}\label{plasdisappoptic}
\omega_+(q)&\approx& \sqrt{qq_{TF}v_F},\\
\omega_-(q)&\approx& qv_F\frac{1+q_{TF}d}
{\sqrt{1+2q_{TF}d} }.\label{plasdisappacoustic}
\end{eqnarray}
\end{mathletters}
in agreement with Ref.\ \onlinecite{sant88}.
Thus the optical mode has the familiar square-root behavior
while the acoustic mode is linear in $q$. Because the latter
is lower in energy, it gives the dominant contribution to the
drag rate at small temperatures, as we show later.

Fig.\ \ref{fig:dis}
shows the dispersion relation of the plasmons as given
by Eq.\ (\ref{plasdis}), for two different values
of layer sepation $d$. Also shown is the part of the
$q-\omega$ plane where particle-hole excitations are
allowed (i.e., Im$[\chi(q,\omega)]\ne 0$) at zero temperature.

In the next section, however, we will see that the plasmon contribution
to the drag rate is in fact given by $q$-values in the intermediate
range, and thus in order to get a quantitative theory of
the collective mode enhancement, use of the small $q$ expansion is
not sufficient, and it is necessary to perform a full numerical
calculation.

\section{The drag rate for the shifted Fermi-Dirac distribution}
\label{sec:shiftFD}

As mentioned previously, when the dopants are far enough away from the
quantum well ($\gtrsim$ 400~\AA), the distribution functions are to
a very good approximation shifted Fermi-Dirac functions, and one can use
Eq. (\ref{dragrate}) to evaluate $\tau_D^{-1}$.  We do so in this
section.

\subsection{Plasmon-pole approximation}

First, we develop a pole approximation for the plasmon
contribition to the drag rate and compare it with the
full numerical solution.
For frequencies near the zeros of the real part of the
dynamical dielectric function we may approximate
$\epsilon(q,\omega)\equiv \epsilon'+i\epsilon''$ in
Eq.\ (\ref{eps0}) as
\begin{mathletters}
\begin{eqnarray}
|\epsilon(q,\omega)|&\approx&
2V(q)e^{-qd} |\beta_\pm(q) (\omega-\omega_\pm(q))
+i\chi''(q,\omega_\pm)|,\\
\beta_\pm(q)&=&\left.\frac{d\chi'(q,\omega)}{d\omega}\right|_{\omega=\omega_\pm}
\end{eqnarray}
\end{mathletters}

Inserting the approximate dielectric function into
Eq.\ (\ref{dragrate}) and assuming that the imaginary part of
$\chi$ is small, we approximate the Lorenzian by $\delta$-functions
and obtain
\begin{equation}
\left|\frac{U_{12}(q)}{\epsilon(q,\omega)}\right|^2\approx
\frac{\pi}{4\mathrm{Im}[\chi(q,\omega)]|\beta_\pm(q)|}
\delta(\omega-\omega_\pm(q)),
\end{equation}
which leads to the following two plasmon contributions to the drag rate
\begin{equation}
\frac{1}{\tau_\pm}=\frac{\hbar^2}{8\pi en_2m_1kT}
\int_0^{q_{c,\pm}} dq\, q^3
\frac{\mathrm{Im}[\chi(q,\omega_\pm(q))]}
{4|\beta_\pm(q)|\sinh^2(\hbar\omega_\pm(q)\beta/2)} \label{dragplas}
\end{equation}
The parameter $q_c$ in this expression
defines the value of $q$ where the plasmon
ceases to exist.  Operationally, we take $q_{c,\pm}$ to be the wavevector
at which the plasmon dispersions disappear (i.e., there are no more
solutions Re$[\epsilon(q,\omega)] = 0$) in the finite-$T$ RPA formalism.
Interestingly, the exponential dependence on the well separation $d$
has dropped out of the integrand in Eq.\ (\ref{dragplas}).
However, $\tau_\pm$ is still $d$-dependent, through the $d$-dependence
of $\omega_\pm(q)$, $\beta_\pm(q)$ and $q_c$.  As $d$ is increased,
the slope of $\omega_-(q)$ increases and $q_c$ decreases, as
evidenced in Fig.\ \ref{fig:dis} (at finite temperatures, the acoustic
and optic plasmons merge when they come close, and the merging
point gives $q_c$).  Both these effects tend to decrease the
integrand and hence the drag rate.  On the contrary, the decrease of
$\omega_+(q)$ with increasing $d$ tends to make $\tau_D^{-1}$ bigger,
but this is generally a weaker effect.
Hence, as expected, $\tau_D^{-1}$ decreases with increasing
well-separation.  Obtaining a precise analytic functional dependence
of $\tau_D^{-1}$ on $d$ is difficult, but we have shown numerically
that in the plasmon-dominated region $\tau_D^{-1}\sim d^{-\alpha}$
where $\alpha\approx 3$.\cite{flen94}
Also note that Eq.\ (\ref{dragplas}) breaks down for large $d$
because in this limit, $\omega_+(q)$ and $\omega_-(q)$ come so close
together that the Lorenzians from both these lines overlap.
This contradicts the initial assumption that the Lorenzians
individually can be approximated by $\delta$-functions.

{}From Eq. (\ref{dragplas}) we see that the plasmon contribution
is given an integral over, among other factors,
the imaginary part of $\chi$ for frequencies and wave-vector corresponding
to the plasmon dispersions. Therefore a non-zero Im[$\chi$] at
$\omega_{\pm}(q)$, which is outside the $T=0$ particle-hole continuum, is
necessary to obtain any plasmon enhancement effect.  Since
the $T=0$ form of Im[$\chi$] is always zero at $\omega_{\pm}(q)$, it
will never give a plasmon enhancement.  One must therefore use the
finite-temperature form of $\chi$ in the evaluation of the drag rate.

At small temperatures, Im[$\chi$] at $\omega_{\pm}(q)$ is small because
the carriers generally do not have sufficient energy to be excited
very far above the Fermi surface.  However, at intermediate temperatures
(on the scale of the Fermi-temperature) there are enough thermally
excited particles so that Im[$\chi$] at $\omega_{\pm}(q)$ is large
and the plasmons actually dominate the drag response.

In Fig.\ \ref{fig:integrands} we show the integrand of the
Eq.\ (\ref{dragrate}) which gives the drag rate at
two different temperatures (for layers of zero width).
There are pronounced peaks are caused by the
presence of plasmons due to the dynamically screened
interlayer interaction.
We also show the plasmon-pole approximation to the integrand
described above, which indicates that this approximation is very good
for temperatures less than $0.5 T_F$ for the relevant $q$'s which
contribute significantly to the integral.

In Fig.\ \ref{fig:contours} we show a contour plot of the
integrand of Eq. (\ref{dragrate}) for $T=0.3T_F$.  Note that
the integrand has significant weight at intermediate $q$ and $\omega$
values, i.e. up to $q\sim .8 k_F$ and $\hbar\omega\sim 2E_F$,
showing that small $\omega$- and $q$-expansions are generally
inadequate for calculating $\tau_D^{-1}$ at intermediate and high
temperatures.


\subsection{Numerical evaluation: $\tau_D^{-1}$}
\label{sec:res}

We evaluated $\tau_D^{-1}$ for parameters corresponding to high
mobility GaAs samples similar to those of ref.\ \onlinecite{gram91}.
We did this several different ways.
First, we numerically integrated Eq. (\ref{dragrate}), using
the finite-temperature RPA Im$[\chi]$ obtained through the technique described
in Appendix \ref{app:X}.  Fig.\ \ref{fig:plas} shows the result of
this full numerical integration of Eq.\ (\ref{dragrate}) (solid line).
We also solved for the temperature dependent plasmon poles numerically and
used the plasmon-pole approximation described above to obtain
plasmon contribution to the drag rate.  One can see that the approximate
curves reproduce to a certain degree the enhancement seen in the full solution.
At low temperatures the upturn in
the drag rate is caused mainly by the acoustic mode, which is lower
in energy and hence easier to excite thermally, while at
higher temperatures both modes contribute. At even higher temperatures
($T\gtrsim 0.6 T_{\mathrm{F}}$) the separation into plasmon and
non-plasmon contributions is not well-defined
because the plasmons are heavily Landau damped and merge into the
single-particle excitation continuum; cf.,
Fig.\ \ref{fig:integrands}.  The Landau damping weakens the
plasmon enhancement effect, causing the scaled $\tau_D^{-1}$ to
peak at around $T \approx 0.5 T_F$.  At extremely high temperatures,
$\tau_D^{-1}\sim T^{-3/2}$, as shown in Appendix \ref{app:largeT}.

Fig.\ \ref{fig:plas} also shows the result of previously
used expressions for the screened interaction. The
dashed line represents the approximation that
the temperature is set to zero in the polarization function used
by Zheng and MacDonald.\cite{zhen93}  As discussed earlier, this
approximation misses the plasmon contribution, as does
the static screened approximation used by Jauho and Smith,\cite{jauh93}
given by the dotted curve.
Neither of these curves shows the plasmon enhancement
demonstrated by the full numerical evaluation of Eq. (\ref{dragrate}).

We have also evaluated of $\tau_D^{-1}$ for densities which are smaller and
larger than the one used in the experiment of Ref.\ \onlinecite{gram91}.
Calculations relevant for that particular density are shown in
Ref.\ \onlinecite{flen94}.
In Fig.\ \ref{fig:res} we show the
scaled drag rate as a function of the temperature.
In both plots we have used a well separation of 800~\AA.
The full curves are for matched densities, while for the dashed (dotted)
the density of layer 2 is twice (half) that of layer 1.
In  Fig.\ \ref{fig:res}(a) the layer 1 density is
5. $\times$ 10$^{11}$ cm$^{-2}$, and in (b) it
is 1. $\times$ 10$^{11}$ cm$^{-2}$.
Since the Fermi temperature is thus 5 times higher in (a),
the peak due to the plasmon enhancement occurs at a much higher temperature.
Also note that the large peaks in the
scaled drag rate occur only in the $n_2/n_1 = 1$ curves.

The insets show the drag rate as a function of the relative
density ratio, which show clear peaks at matched densities.
The peaks are signatures of the plasmon enhancement effect,
and they occur when the Fermi {\em velocities} $v_{F,i}$
of both subsystems are equal.\cite{flen94}
This is because the size of the plasmon contribution to $\tau_D^{-1}$
is determined by the smaller of the Im$[\chi_i(q,\omega_{\pm}(q))]$,
which is roughly given by the distance between the particle-hole
continuum and the plasmon dispersion line on the $\omega-q$ plane.
The min\{Im$[\chi_i]$\} is maximized at matched $v_{F,i}$.
As the $v_{F,i}$'s start to differ from each other, the particle--hole
continuum of the subsystem with the smaller
$v_F$, say subsystem 2, starts to move away from the plasmon dispersion
lines (which lie above the particle-hole continuum of the subsystem 1
with the larger $v_F,1$).  Consequently, the Im$[\chi_2(q,\omega_\pm(q))]$
of the subsystem 2 is decreased with increasing mismatch of $v_F$,
which accordingly decreases the total plasmon contribution to the drag
rate.  The peak at equal Fermi velocities (and hence matched densities
when the masses of the wells are equal) should be the feature of
the predicted plasmon enhancement which is easiest
to experimentally verify.
Other signatures include the aforementioned $d^{-3}$ behavior and
the upturn in the temperature scan at approximately
$0.2 T_F$.\cite{flen94}

\section{Drag rate as a function of dopant distance from quantum well}
\label{sec:tauvary}

Thus far, we have assumed that the $\tau(k)$ is constant (which is
equivalent to assuming that the distribution function is a shifted
Fermi-Dirac function) and hence Eq.\ (\ref{dragrate}) is valid.  This is only
true if the sum of the impurity and phonon scattering rates is
relatively energy independent and/or these rates are much smaller than
the intralayer electron--electron scattering rates, as is the case
when the dopants are placed approximately 700~\AA\ away from the
quantum well.  We find, however, than when the impurities are moved
closer into the wells, Eq.\ (\ref{dragrate}) is no longer valid;
in fact, when the impurities are placed 100~\AA\ from the side of the well,
it underestimates the drag rate by approximately a factor of 2 when
compared to Eq. (\ref{tauD}) because it neglects the energy dependence
of the momentum relaxation time $\tau(\mbox{\boldmath{$k$}})$.

To evaluate the drag rate in this case, one must be able to calculate
the momentum relaxation rate $\tau(\mbox{\boldmath{$k$}})$ in the presence of
impurity, phonon and electron--electron scattering.
We adapt the formalism used to study this in
3-dimensions\cite{sanb95} to 2-dimensions.\cite{hu95}
We assume that the impurities are charged and are in an uncorrelated
$\delta$-doped layer a distance $s$ from the side of the quantum
well.  We calculate $\tau(k)$ including effects of the dynamically
screened intralayer electron--electron, screened remote
ionized impurity and screened acoustic
(both deformation potential and piezoelectric) phonon\cite{kawa92} scattering.
Our calculations indicate that the $\tau(k)$ is exceedingly flat
when $s = $ 700\AA, but as the impurities are moved in $\tau(k)$ starts
to show a significant {\em positive} slope at around $s = $ 300 \AA.
This is because the charged impurity potential is long-ranged,
implying that it falls off rapidly with increasing momentum
transfer,\cite{ridley} and hence
the momentum relaxation time increases with increasing momentum.
At large $s$, the electron--electron scattering, which drives the
distribution to a shifted Fermi-Dirac function, suppresses this positive
slope.  However, as the impurities are moved closer to the well,
the impurity scattering and electron--electron scattering rates
become comparable, and hence the positive slope in $\tau(k)$ develops.

A positive slope in $\tau(k)$ enhances the drag rate
because it implies that there are relatively more carriers
at higher energies (see Eq. (\ref{definetau})).  These higher
energy carriers are able to carry more current, which consequently
causes increase in the drag rate.  The enhancement is even more
pronounced when we include dynamical screening because it is
the high-energy particles which participate in the
plasmon-mediated interlayer Coulomb interaction.
In Fig. \ref{fig:tauvary} we show the drag rate
as a function of the charged impurities from the side of the wells
$s$ (which we assume is the same for both wells), for several
different temperatures, all other parameters being fixed.  The figure
shows that one can obtain an enhancement in the drag rate of
approximately a factor of 2 by moving the impurities to within
100~\AA\ of the side of the well, all this coming from the momentum
dependence of $\tau(k)$.  From the figure, one can see the effect
of the plasmons again on $\tau_D^{-1}$ in the fact that
the enhancement at small $s$ is largest at $T=30$K where the plasmon
enhancement is the greatest, because a larger number of high-energy
particles can now partake in the plasmon-enhanced interlayer scattering.

Finally, since we have studied the drag rate in the case where
the absolute magnitude of the impurity scattering rate is large
(because $s$ is small) one could ask if
the diffusive form of the polarizability
function\cite{zhen93,flen95a,kame95}
at small $q$ and $\omega$ has any effect on our calculations
at experimentally relevant temperatures.  We conclude it does not
because the crossover temperature is below which the diffusive
effect should be seen goes as\cite{zhen93}
$\sim 0.1\mbox{K} \exp[-0.9(\ell/d)^2]$.  Since
$d$ is on the order of 400~\AA, and $\ell$, the mean free path,
is approximately 0.1 to 1 $\mu$m for $s$ = 100\AA, the crossover
temperature is extremely small and experimentally irrelevant.

\section{Discussion and summary}
\label{sec:disc}

Using the Boltzmann equation, we have derived a formula for the
transresistivity of coupled two-dimensional electron gases,
valid for arbitrary intralayer scattering,
to lowest nonvanishing order in the interlayer interaction.
While the Boltzmann equation does not include quantum effects such as
weak-localization, it has been shown\cite{kame95,flen95a} that these
effects for systems where $k_F \ell \gg 1$ are negligible.
The transresistivity depends on the momentum relaxation times
$\tau_i(\mbox{\boldmath{$k$}})$ of both layers, which are given by the
linear-response
solution of the distribution function to an applied electric field).

For isotropic parabolic bands, one can define a drag rate which
is proportional to the transresistivity.  In this case, in the limit
where $\tau(k)$ is a constant, one regains previously the obtained result,
Eq. (\ref{dragrate}), from our result, Eq. (\ref{tauD}).
A constant $\tau(k)$ is equivalent to a shifted Fermi-Dirac
distribution function under application of a small electric field.
In experiments done previously, where the dopant layer
in the modulation-doped structures is relatively far away from the
quantum wells, $\tau(k)$ is in fact relatively constant over the
relevant energy range (i.e., within a few $k_B T$ from the chemical
potential), because the electron--electron scattering rate, which
tends to drive the distribution to a shifted Fermi-Dirac function,
dominates over all other scattering rates.  Thus, for calculations involving
these structures, we have used Eq. (\ref{dragrate}).

The interlayer coupling is given by the screened Coulomb interaction.
Due to the presence of plasmons, both acoustic and optic, the
interaction can be significantly enhanced, as these plasmons serve
to ``anti-screen" the interaction.   At low temperatures, the
plasmons are ``frozen out," and hence they play no role.  At higher
temperatures, however, the plasmons can enhance the drag rate by
almost an order of magnitude over interactions which exclude
dynamic effects of the coupled electron gas system.
We have shown that the maximum
plasmon enhancement occurs around 30 K for GaAs when both wells are
doped at 1.5$\times 10^{11} \mbox{cm}^{-2}$.  Furthermore, because
the plasmons enhancement is most effective when the Fermi velocities
of the two electron gases are equal, one should see a peak in the
$\tau_D^{-1}$ as the ratio of densities of the electron gases $n_2/n_1$
is varied through 1.

We have also investigated the effect of moving impurities closer to
the quantum wells, which gives $\tau(k)$ a positive slope.  This
increases the number of high-energy (and hence large current-carrying)
particles and hence also enhances the drag rate.  The effect is largest
when the the plasmons effect is the greatest.  One obtains an
enhancement of approximately a factor of 2 when impurities are moved
in from $s = $ 400~\AA\ to $s =$ 100~\AA.

\section*{Acknowledgment}

We thank A.-P. Jauho for numerous discussions.
KF was supported by the Carlsberg Foundation.

\appendix

\section{Boltzman equation for single layer}
\label{app:boltz}

In this Appendix, we describe the formalism for the Boltzmann equation
in a single quantum well, which we use in the main text.

\subsection{Formalism for intralayer scattering}

The linearized form of the Boltzmann equation for (positively charged)
carriers with an electric field {\boldmath $E$} is
\begin{equation}
e\mbox{\boldmath{$E$}}\cdot\Bigl({\partial{f^0}\over\partial{\mbox{\boldmath{$p$}}}}\Bigr) \equiv
e\mbox{\boldmath{$E$}}\cdot\mbox{\boldmath{$v$}}\Bigl({\partial{f^0}\over\partial{\varepsilon}}\Bigr)
= -\hat H[\psi(\mbox{\boldmath{$k$}})].
\label{boltzeq}
\end{equation}
{}From the principle of detailed balance, one can show that
$\hat H$ is a Hermitian operator\cite{smith}; i.e.,
for arbitrary functions $a(\mbox{\boldmath{$k$}})$ and
$b(\mbox{\boldmath{$k$}})$
\begin{equation}
\int d\mbox{\boldmath{$k$}}\; a(\mbox{\boldmath{$k$}})\; \hat
H[b](\mbox{\boldmath{$k$}})
= \int d\mbox{\boldmath{$k$}}\; b(\mbox{\boldmath{$k$}})\; \hat
H[a](\mbox{\boldmath{$k$}}).
\end{equation}

Therefore, the current in the $\beta$-direction
is given by Eq.\ (\ref{definepsi}) and (\ref{boltzeq}),
\begin{eqnarray}
j_\beta &=& 2 e\int \frac{d\mbox{\boldmath{$k$}}}{(2\pi)^d}\
v_\beta\ \delta\! f(\mbox{\boldmath{$k$}}) \nonumber\\
&=& - 2e  k_B T \int \frac{d\mbox{\boldmath{$k$}}}{(2\pi)^d}\
v_\beta(\mbox{\boldmath{$k$}})
\ \Bigl({\partial{f^0}\over\partial{\varepsilon}}\Bigr)\
\psi(\mbox{\boldmath{$k$}})\nonumber\\
&=& 2 k_B T e^2 \sum_\alpha E_\alpha
\int \frac{d\mbox{\boldmath{$k$}}}{(2\pi)^d}\ v_\beta(\mbox{\boldmath{$k$}})
\ \Bigl({\partial{f^0}\over\partial{\varepsilon}}\Bigr)(\mbox{\boldmath{$k$}})\
\hat H^{-1} \left[v_\alpha
\Bigl({\partial{f^0}\over\partial{\varepsilon}}\Bigr)\right](\mbox{\boldmath{$k$}})\nonumber\\
&\equiv& n e \sum_{\alpha}\mu^{\beta\alpha}_t E_\alpha,
\label{jbeta}
\end{eqnarray}
where the factor of $2$ is for spin, and $\mu_t^{\beta\alpha}$
is mobility tensor of the system.

\subsection{Impurity scattering}

We now examine a concrete example, an isotropic system dominated by
impurity scattering, in which case the linearized collision operator
is given by
\begin{eqnarray}
-\hat H[\psi](\mbox{\boldmath{$k$}}) =
k_B T\Bigl({\partial{f^0}\over\partial{\varepsilon}}\Bigr)(k)\, \tau^{-1}(k)\,
\psi(\mbox{\boldmath{$k$}}),
\end{eqnarray}
where $\tau(k)$ is the {\em transport} lifetime.\cite{smith}
The inverse of $\hat H$ is trivial,
\begin{equation}
\hat H^{-1}[g](\mbox{\boldmath{$k$}}) = -\frac{g(\mbox{\boldmath{$k$}})
\tau(\mbox{\boldmath{$k$}}) }{k_B T\
(\partial f^0/\partial \varepsilon)(\mbox{\boldmath{$k$}})},
\end{equation}
and hence the solution of Eq.\ (\ref{boltzeq}), for $\mbox{\boldmath{$E$}}$ in
the
$x$-direction, is
\begin{equation}
\psi(\mbox{\boldmath{$k$}})= \frac{e E\,
v_x(\mbox{\boldmath{$k$}})\,\tau(k)}{k_B T}.
\end{equation}

\section{Numerical evaluation of the finite temperature polarization function}
\label{app:X}

\title{Numerically efficient methods for calculating the real and
imaginary parts of the 2-dimensional finite-temperature RPA polarization
function}

Here we describe an efficient technique for calculating the
finite-temperature 2D-RPA-$\chi$ for an isotropic parabolic band.
The RPA $\chi$ is given by
\begin{equation}
\chi(\mbox{\boldmath{$q$}},\omega) = \frac{2m}{\hbar^2}\int
\frac{d\mbox{\boldmath{$k$}}}{(2\pi)^2}
\frac{f^0(\mbox{\boldmath{$k$}}+\mbox{\boldmath{$q$}}/2) -
f^0(\mbox{\boldmath{$k$}}-\mbox{\boldmath{$q$}}/2)}
{\mbox{\boldmath{$k$}}\cdot\mbox{\boldmath{$q$}} - m\omega/\hbar - i 0^+},
\end{equation}
where $f^0$ is the Fermi distribution.
We define the following nondimensional quantities
\begin{equation}
Q = q/k_F,\quad
K = k/k_F,\quad
\Omega = \hbar\omega/E_F,\quad t = k_B T/E_F,\quad
\tilde \mu = \mu/E_F,\quad
\tilde \chi = \frac{\chi}{m/(\pi \hbar^2)},
\end{equation}
where $k_F = (2\pi n )^{1/2}$ ($n$ is the density), and
$E_F = \hbar^2 k_F^2 / (2m)$.  Note that $\tilde \chi$ is
normalized by the value of $\chi(q\rightarrow 0,\omega=0; T=0)$.
In these units,
\begin{equation}
\tilde\chi(Q,\Omega) = \frac{1}{2\pi}
\int dK_x\, dK_y\;
\frac{f^0(K_x + Q/2, K_y) - f^0(K_x - Q/2, K_y)}
{ K_x Q - \Omega/2 - i0^+},
\end{equation}
where
\begin{equation}
f^0(K_x,K_y) = \frac{1}{\exp([K_x^2 + K_y^2 - \tilde\mu]/t) + 1}; \quad
\tilde \mu =  t\; \ln (e^{1/t} - 1).
\end{equation}

\subsection{Imaginary part of $\chi$}

The imaginary part of $\chi$ is given by
\begin{eqnarray}
\mathrm{Im}[\tilde\chi(Q,\Omega)] &=&
\frac{1}{2 Q} \int_{-\infty}^\infty dK_y\
f^0(\frac{\Omega}{2Q} + \frac{Q}{2}, K_y) -
f^0(\frac{\Omega}{2Q} - \frac{Q}{2}, K_y)\nonumber\\
&=& \frac{\sqrt{t\pi}}{2 Q}
\left[ {\cal F}_{-1/2} (\frac{A_+}{t}) - {\cal F}_{-1/2}(\frac{A_-}{t})
\right];\quad
A_{\pm} = \tilde \mu - \left( \frac{\Omega}{2Q} \pm \frac{Q}{2}
\right)^2,
\end{eqnarray}
where
\begin{equation}
{\cal F}_{-1/2} (x) = \frac{1}{\sqrt{\pi}}
\int_0^\infty dy \
\frac{y^{-1/2}}
{\exp(y - x) + 1},
\end{equation}
is the Fermi function of order $-1/2$.\cite{blake}
${\cal F}_{-1/2}$ is a well-known function, and many excellent
approximation schemes are available to evaluate it.\cite{cody}

\subsection{Real part of $\chi$}

We use the expression by Maldague\cite{mald78}
\begin{equation}
\chi(q,\omega;\mu,T) = \int_0^\infty d\mu'\ \chi(q,\omega;\mu',T=0)
\frac{1}{4 k_B T \cosh^2[(\mu-\mu')/(2k_B T)]}.
\end{equation}
Letting $a_{\pm} =\frac{1}{2}(\frac{\Omega}{Q} \pm Q)$ and using
the known form for $\chi(q,\omega;T=0)$, we obtain
\begin{eqnarray}
\mathrm{Re}[\tilde\chi(Q,\Omega)]
&=& - \int_0^\infty d\tilde\mu' \ \frac{1}{4t\cosh^2[(\tilde\mu-\tilde\mu')
/(2t)]}
\Bigl[ 1 - \frac{\mathrm{sgn}(a_+)}{Q}\;\sqrt{a_+^2 - \tilde\mu'}
\;\theta( a_+^2 - \tilde\mu') +  \nonumber\\
&& \qquad \frac{\mathrm{sgn}(a_-)}{Q}\;
\sqrt{a_-^2 - \tilde\mu'}\;
\theta(a_-^2 - \tilde\mu')
\Bigr] \nonumber\\
&=& - \frac{1}{\exp(-\tilde\mu/t) + 1}
+ \frac{\mathrm{sgn}(a_+)}{Q}\; M_t(a_+^2)
- \frac{\mathrm{sgn}(a_-)}{Q}\; M_t(a_-^2)
\label{repi}
\end{eqnarray}
where
\begin{equation}
M_t(x) = \int_0^{x} d\tilde\mu' \frac{1}{4 t \cosh^2(
[\tilde\mu'- \tilde\mu(t)]/(2t))} \left(x -
\mu'\right)^{1/2}.
\end{equation}
For any given temperature, this function $M_t(x)$ need only be evaluated
once, and then a polynomial, Pad\'e or a spline fit can be made to it.
Then, Re$[\chi(q,\omega)]$ can be evaluated with this fitted function,
using Eq.\ (\ref{repi}).
Caution, however, should be exercised when $Q \ll 1$ and
$\Omega/Q$ is not small.  Then,
the terms $\pm Q^{-1}M_t(a_{\pm})$ in Eq.\ (\ref{repi}) are large
and approximately cancel, leading to large numerical uncertainly.
We got around this problem by
using the well-known asymptotic form Re$[\tilde\chi(Q,\Omega)] =
2 Q^2/\Omega^2$, which can be derived from Eq.\ (\ref{repi}).
This form should be used in the regions when numerical inaccuracies plague
the form given in Eq.\ (\ref{repi}).

\section{The large temperature limit $T\gg T_{\mathrm{F}}$}
\label{app:largeT}
\narrowtext

In the case of very large temperature, the Coulomb interactions are
basically unscreened because the typical wavevector transfers is much larger
than the debye screening wavenumber, $q_D = 2\pi e^2 n/(\epsilon_o
k_B T)$, and one can assume that the interlayer interaction is
the bare unscreened interaction.
Then, under further assumptions that the carriers are
nondegenerate and the intralayer momentum scattering rate is
constant, the we show that the drag rate goes as
$\tau_D^{-1} \propto
T^{-3/2}$.  Since $\tau_D^{-1} \propto T^2$ for small $T$,
the drag rate must peak at some temperature.

The polarization function for a Maxwell Boltzmann distribution in the
RPA is
\begin{equation}
\tilde\chi(\tilde q,\tilde\omega) =
\frac{1}{2t\tilde q}
\left[Z(\frac{\tilde\omega}{2\tilde q} + \frac{\tilde q}{2})
-     Z(\frac{\tilde\omega}{2\tilde q} - \frac{\tilde q}{2})\right],
\end{equation}
where $Z$ is the plasma dispersion function.\cite{fried}
Here as before, $\tilde\chi = \chi/[m/(\hbar^2 \pi)]$ and $t = T/E_F$,
whereas
the other variables with tildes are defined as being normalised
with respect to temperature,
\begin{eqnarray}
\tilde q &=& \hbar q/\sqrt{2 m k_B T}, \nonumber\\
\tilde\omega &=& \frac{\hbar\omega}{k_B T},\nonumber\\
\tilde d &=& d \sqrt{2 m k_B T} \hbar^{-1}.
\end{eqnarray}
The classical limit ($\tilde q \ll 1$) of the polarization function
$\chi_{\mathrm{cl}}$ [which is valid here because the $\tilde q$
integration is cut off at small $\tilde q$; see below]
only depends on the ratio $\tilde\omega/\tilde q$,
\begin{equation}
\tilde\chi_{\mathrm{cl}}(\tilde\omega/\tilde q)
= \frac{1}{2t}\; Z'\left(\frac{\tilde\omega}{2\tilde q}\right).
\end{equation}
The imaginary part is given by\cite{fried}
\begin{equation}
\mathrm{Im}[\chi_{\mathrm{cl}}(\tilde\omega/\tilde q)]
= - \frac{\tilde \omega \sqrt{\pi}}{2 t \tilde q} \exp\left(-\frac{
\tilde\omega^2}{4\tilde q^2 }\right).
\end{equation}

{}From Eq.\ (\ref{dragrate}), the drag rate for identical layers
is
\begin{eqnarray}
\tau_D^{-1} &=& \left\{
\frac{\hbar q_F^4}{8\pi^2 n m} \left(\frac{2\pi e^2 m}{\epsilon_0
\pi \hbar^2 q_F}\right)^2 \right\}\ \
t \int_0^\infty d\tilde q\ \tilde q^3 \nonumber\\
&&\times
\int_0^\infty d\tilde\omega \ \frac{\exp(-2 \tilde q \tilde d)}
{\tilde q^2 \sinh^2(\tilde\omega/2)}\
\frac{\pi\tilde\omega^2}{4 t^2\tilde q^2} \exp\left(-\frac{
\tilde\omega^2}{2\tilde q^2}\right).
\end{eqnarray}
Having nonidentical layers simply leads to a change in the prefactor.
Defining $\tilde\tau_D^{-1}$ as $\tau_D^{-1}$ divided by the
non-temperature-dependent prefactor in the curly braces in the
above equation yields
\begin{equation}
\tilde\tau_D^{-1} = \frac{\pi}{4 t}
\int_0^\infty d\tilde q\ \tilde q^{-1} \exp(-2 \tilde q \tilde d)
\times \int_0^\infty d\tilde\omega \frac{\tilde\omega^2 \exp(-\tilde
\omega^2/(2\tilde q^2)) } {\sinh^2(\tilde\omega^2/2)}.
\end{equation}

Since $\tilde d \propto T^{1/2}$, for large $T$
the integrals are cut off at small
$\tilde q$.  The presence of the gaussian term implies in the
$\tilde\omega$ integral implies that this integral is also cut off at
small values, and
hence an expansion $\sinh^2(\tilde\omega/2) \approx \tilde\omega^2/4$
can be made.  This gives
\begin{eqnarray}
\tilde\tau_D^{-1} &\approx& \pi t^{-1} \int_0^\infty d\tilde q\;
\tilde q^{-1} \int_0^\infty d\tilde\omega\; \exp(-\tilde\omega^2/
(2\tilde q^2)), \nonumber\\
&=& \frac{\pi}{t} \sqrt{\frac{\pi}{2}} \int_0^\infty d\tilde q\;
\exp(-2 \tilde q \tilde d),\nonumber\\
&=& \left(\frac{\pi}{2}\right)^{3/2} \frac{1}{t\tilde d}.
\end{eqnarray}
Since $t \propto T$ and $\tilde d \propto T^{1/2}$, this
shows that the temperature dependence of the drag rate in the
large temperature limit, given an energy-independent intralayer
momentum relaxation rate, is
\begin{equation}
\tau_D^{-1} \propto T^{-3/2}.
\end{equation}

In actual fact, this asymptotic behavior is reached very slowly.
Our numerical evaluation of the drag rate indicates that
for the experimental parameters of Ref. \onlinecite{gram91},
this behavior of $\tau_D^{-1}$ occurs only when $T \gg 10 T_F$, and
hence is not experimentally observable for these parameters.



\begin{figure}
\caption{Plasmon dispersions at zero temperature for
two identical wells based on Eq.\ (\protect\ref{plasdis}).
The dispersions are shown for two different values of the layer
separation $k_Fd=4$ (solid lines) and $k_Fd=12$ (dashed lines), and
the hatched area is the particle-hole continuum.
The lower-(upper-) lying branch correspond to charge density
oscillations in the two layers being out of (in) phase.
The modes are well-defined only at small $q$.  At larger $q$, they
disappear either by merging together or due to Landau damping.
\label{fig:dis}
}
\end{figure}

\begin{figure}
\caption{
Scattering rates [integrand of Eq.\ (\protect\ref{dragrate})] as a
function of energy transfer $\hbar\omega$, for GaAs coupled quantum wells with
equal electron densities of $n = 1.5\times 10^{11}\,\mbox{cm}^{-2}$,
well separation $d = 8\, k_F^{-1} = 800$ \AA, $q = 0.3\,k_F$, and zero well
widths, at two different temperatures $T=$ 0.5 $T_F$ and 0.8 $T_F$,
where $T_F = 61$ K.
The dashed lines are the plasmon-pole approximation to the plasmon
peaks, which is used for calculation the pole contributions
in Fig.\ \protect\ref{fig:plas}.  Note how the poles
merge and are Landau damped at higher temperatures, implying that
the plasmon-pole approximation is not valid for $T \protect\gtrsim 0.6 T_F$
and higher.
}
\label{fig:integrands}
\end{figure}

\begin{figure}
\caption{Contour of the integrand of Eq.\ (\protect\ref{dragrate})
for the case of two identical GaAs quantum wells with densities
$n = 1.5\times 10^{11} \mbox{cm}^{-2}$, and a well separation of
of  $d = 3.75\, k_F^{-1} = 375$ \AA, well widths of 200~\AA, as in
the experiment by Gramila {\sl et al.}\protect\cite{gram91},
at $T = 18$ K (=0.3 $T_F$).  Adjacent contours differ by
$10^{-4} E_F/\hbar = 8\times 10^{8}\mbox{s}^{-1}$, and the
rightmost contour line equal to $10^{-4} E_F/\hbar$.
The dashed lines are the plasmon-pole approximation dispersion curves.
\label{fig:contours}
}
\end{figure}

\begin{figure}
\caption{
Temperature dependence of the drag rate scaled by $T^2$, for the
same parameters as in Fig.\ \protect\ref{fig:contours}.
The full bold curve corresponds to calculations using
the finite-$T$ form of $\chi(q,\omega)$,
the dotted curve to using the $T=0$ form of $\chi$,\protect\cite{zhen93}
and the short-dashed is based on the static screening
approximation.\protect\cite{jauh93}
Also shown are the plasmon-pole approximation estimates for the
acoustic plasmon (ap) and optic plasmon (op)
contributions to the $\tau_D^{-1}$, and the sum of the two (op+ap).
For $T \protect\gtrsim 0.6 T_F$, this approximation becomes less reliable
due to large Landau damping of the modes, and hence we have plotted
the results with dashed lines.
\label{fig:plas}
}
\end{figure}

\begin{figure}
\caption{
The drag rate as a function of temperature, for well separation
$d=800$~\AA\ and well width $200$~\AA, for densities (a) $n_1=5\times
10^{11}\mbox{cm}^{-2}$ and (b) $n_1=1\times 10^{11}\mbox{cm}^{-2}$.
The full, dashed and dotted lines are for $n_2/n_1 = 1,\ 2$ and 0.5,
respectively.
Insets: drag rate as a function of relative density $n_2/n_1$,
when $n_1$ is fixed and $n_2$ is allowed to vary at (a) 80 K and
(b) 20 K.  The peaks at $n_2/n_1 = 1$ are evidence of plasmon
enhanced drag.
}
\label{fig:res}
\end{figure}

\begin{figure}
\caption{
The drag rate as a function of the distance of the charged dopant layer,
$s$, from the side of the well.
The solid (dashed) lines were calculated using full dynamic (static)
screening, and $T =$ 15, 20 and 30~K.
The parameters used were $n=1.5\times 10^{11} \mbox{cm}^{-2}$,
$d = 375$\AA, and well width of 100~\AA in GaAs.
For $s\protect\gtrsim400$~\AA,
the distribution functions are very close to shifted Fermi-Dirac
functions.  As $s$ decreases, the distribution functions deviate
from shifted Fermi-Dirac functions, and the drag rate increases markedly.
}
\label{fig:tauvary}
\end{figure}


\begin{thebibliography}{10}
\bibitem{bandcompl}
We have assumed that the scattering rate only depends on the momentum and
  energy transfered. This is valid when the unit-cell periodic part of the
  Bloch wavefunction of the electron does not vary significantly over the
  region of the Broullioun zone relevant to the problem (see, e.g., Ref.
  \onlinecite{ridley}, chapter 3). When this is not true, e.g., in the case of
  holes in the warped region of the Broullioun zone of GaAs, the expression for
  $\tau_D$ cannot be deconvolved in the manner described further on in this
  paper, and the calculation of $\tau_D$ generally requires the numerical
  evaluation of a six-dimensional integral.
\end{thebibliography}
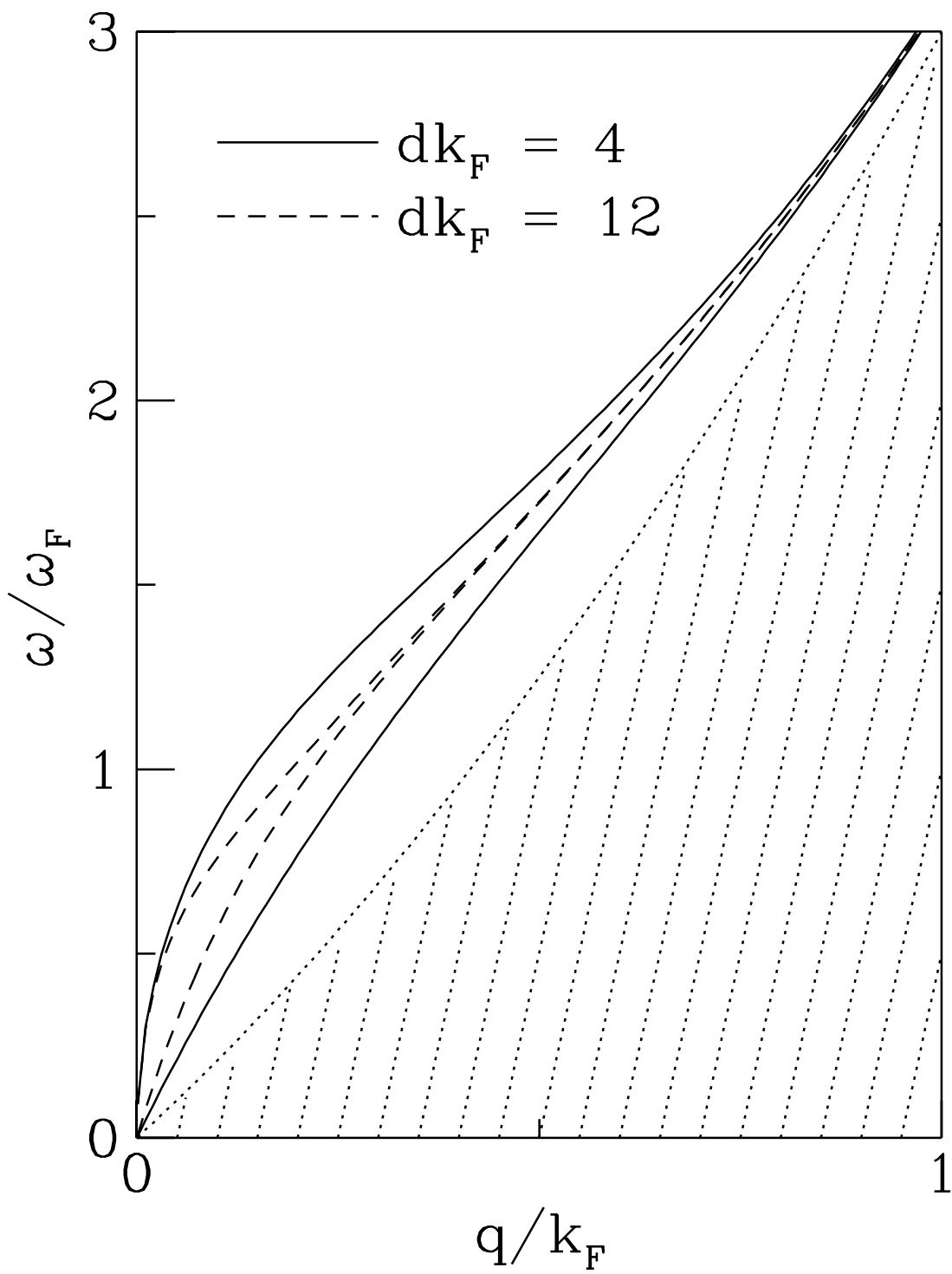
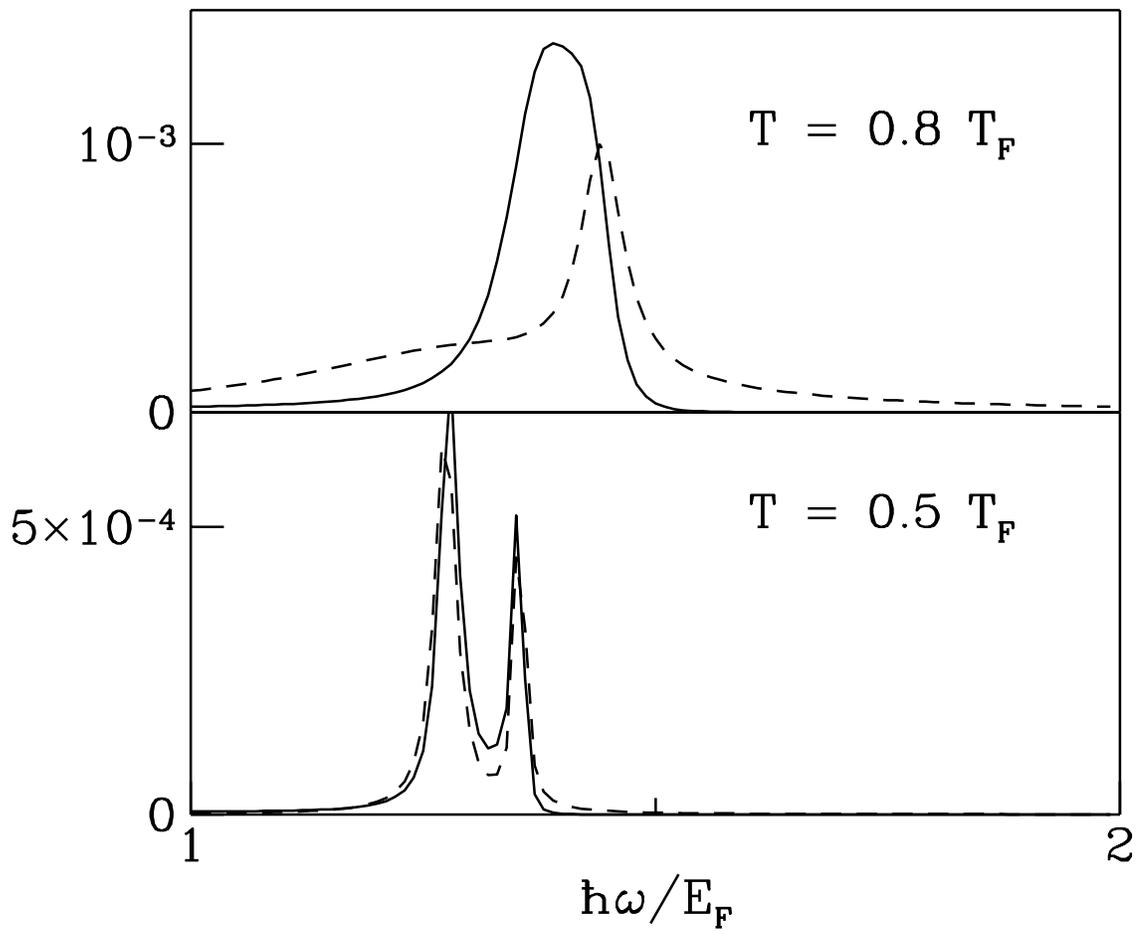
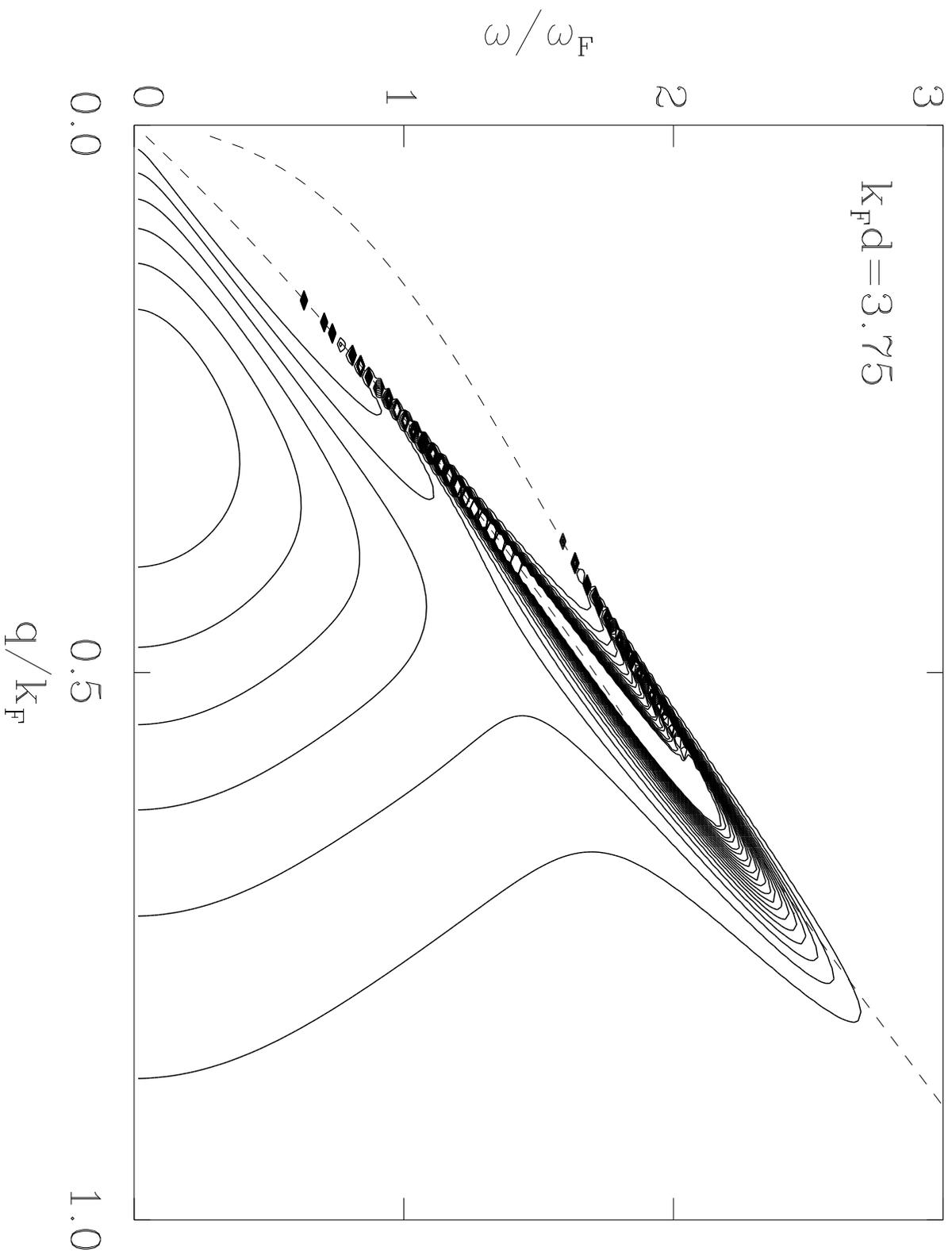
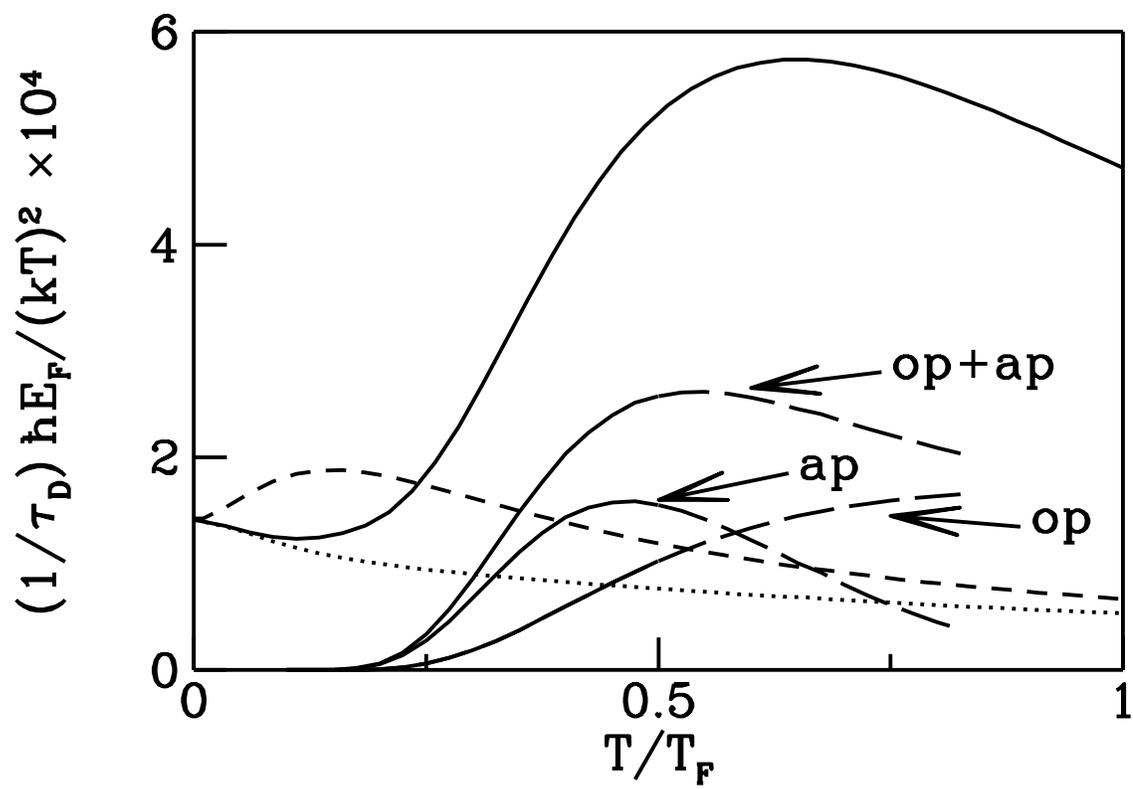
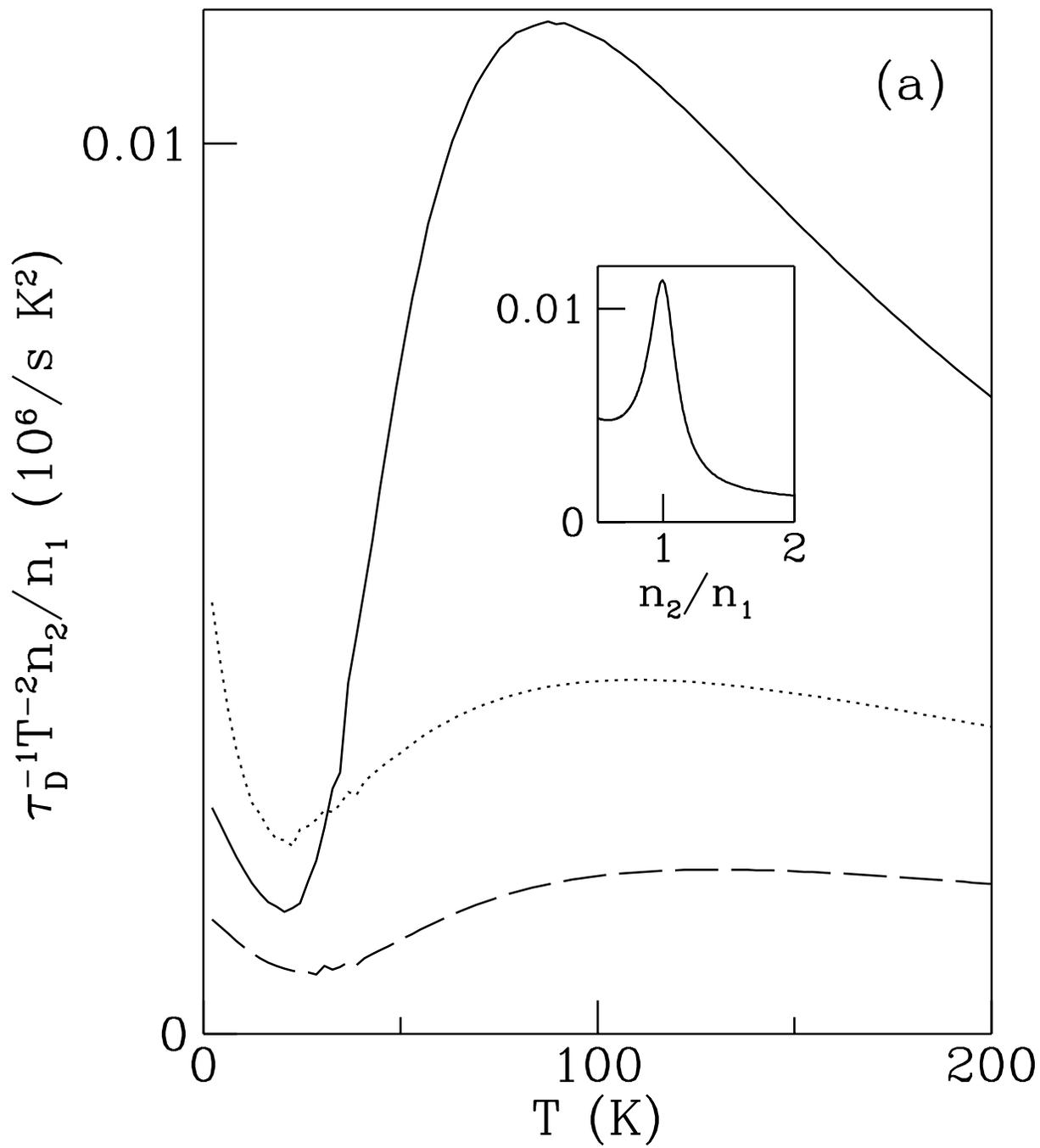
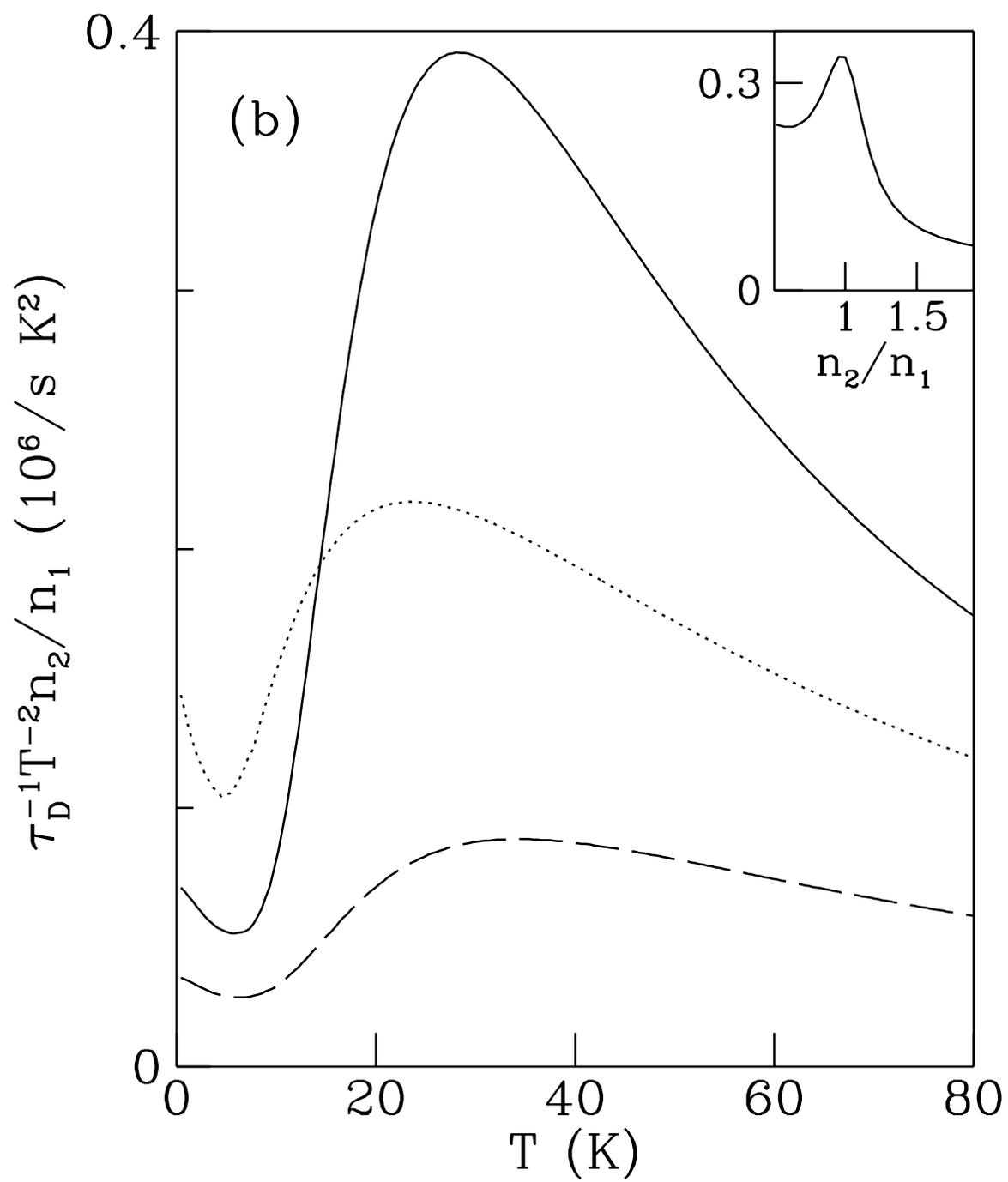
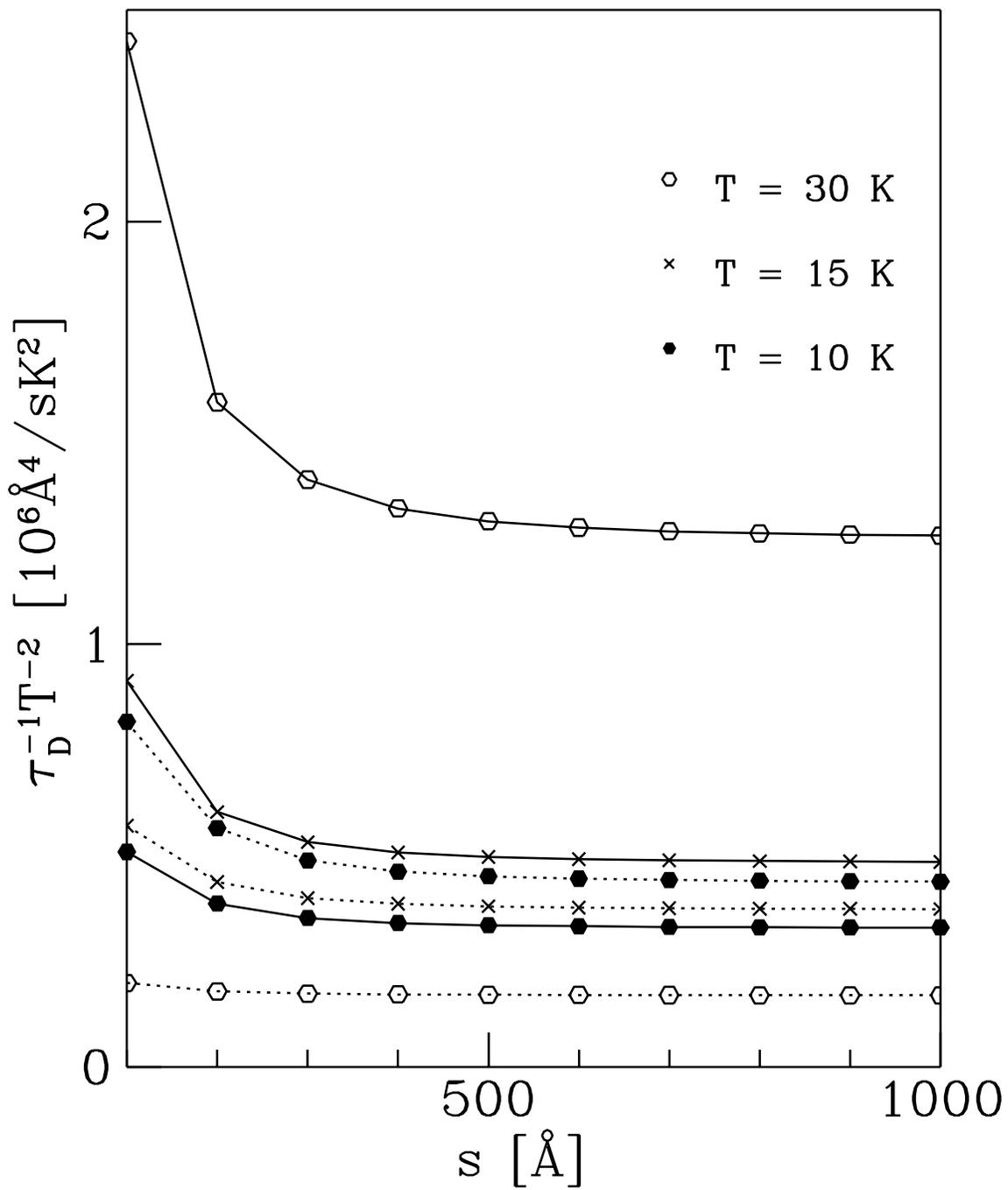
\end{document}